\documentstyle[aps,epsf]{revtex}
\begin{document}
\newcommand{\lapprox}{\stackrel{<}{\scriptstyle \sim}}
\newcommand{\gapprox}{\stackrel{>}{\scriptstyle \sim}}
\title{
\vspace{2cm}
\large
\bf Antiferromagnetic vortex state in a high-temperature superconductor \\
\vspace{0.8cm} }

\author{
B.\ Lake$*$,  
H.\ M.\ R\o nnow$\dag$, 
N.\ B.\ Christensen$\ddag$, 
G.\ Aeppli$\S\ddag$,
K.\ Lefmann$\ddag$, D.\ F.\ McMorrow$\ddag$,
P.\ Vorderwisch$||$, P.\ Smeibidl$||$, 
N.\ Mangkorntong$\P$, T.\ Sasagawa$\P$, M.\ Nohara$\P$, H.\ Takagi$\P$,
T.\ E.\ Mason$\#$. 
}

\address{
$*$Oak Ridge National Laboratory, P.O. Box 2008 MS 6430, Oak Ridge, TN 37831-6430, U.S.A.\\
$\dag$CEA (MDN/SPSMS/DRFMC), 17 Av.\ des Martyrs, 38054 Grenoble cedex 9, France.\\
$\ddag$Materials Research Department, Ris\o\ National Laboratory, 4000 Roskilde, Denmark.\\
$\S$NEC Research Institute, 4 Independence Way, Princeton, New Jersey 08540-6634, USA.\\
$||$BENSC, Hahn-Meitner Institut, Glienicker Strasse 100, 14109 Berlin, Germany.\\
$\P$Department of Advanced Materials Science, Graduate School of Frontier Sciences, University of Tokyo, Hongo 7-3-1,
Bunkyo-ku, Tokyo 113-8656, Japan.\\
$\#$Experimental Facilities Division, Spallation Neutron Source, 701 Scarboro Road, Oak Ridge, TN 37830, U.S.A.\\ 
}

\maketitle

\date{\today}


Magnetic fields penetrate conventional superconductors via normal state 
metallic inclusions, or vortices, which can exist in fluid or frozen 
states. At the macroscopic level, the phenomenology of high-temperature 
(high-$T_{c}$) superconductors follows this paradigm. However, at the 
microscopic level, vortices in high-temperature cuprates differ from those 
in conventional superconductors. Scanning tunneling microscopy indicates 
the persistence of an energy gap for electronic excitations in the vortex 
cores \cite{electronic}, and there are also predictions that the cores are 
magnetically ordered \cite{arovas}. Recent experiments on 
La$_{2-x}$Sr$_x$CuO$_4$ at optimal doping indicate that the individual 
vortices should be considered as nanomagnets with enhanced low frequency 
fluctuations but no static order \cite{science-us}. Here we show that static, 
long-ranged, antiferromagnetism is associated with the vortex state in an 
underdoped sample. 

La$_{2}$CuO$_4$ is an insulator with static commensurate antiferromagnetism. 
For $x$$>$0.05, the magnetic correlations are characterized by incommensurate 
(IC) wavevectors of the type shown in the inset of fig.\ 1b \cite{cheong}. 
Dynamic IC correlations are common to all metallic samples, and for 
$x$$\simeq$1/8, they slow down to yield static order with a range in excess 
of 50 Cu-Cu spacings \cite{tranquada,order1,order2}. Our samples have 
$x$=0.10 and develop superconductivity below a zero-field transition 
temperature of $T_{c}$($H$=0)=29K. The magnetotransport data in fig.\ 1a show 
that even a modest external magnetic field ($H$) strongly suppresses the 
onset of the zero-resistance state. Figure 1b displays the temperature 
($T$)-dependence of the IC order,  established via neutron diffraction. For 
$H$=0, the IC order is weak and $T-$independent up to at least 50K. A field 
of 5T produces a $T$-independent increase of $\sim$50\% in the signal above 
$T_{c}$($H$=0), similar to the enhancement reported at $T$=4.2K and 
$H$=10T for an $x$=0.12 sample with $T_{c}$($H$=0)=12K~\cite{katano}. Our new 
finding is that between $T_{c}$($H$=0) and the irreversibility temperature 
$T_{\rm irr}$($H$=5T), a much stronger field-induced signal appears, 
saturating at over three times the zero-field signal. 

A key to understanding our result is that the measured in-plane magnetic 
correlation length $\zeta$$\ge$600\AA\, is greater than both the 
superconducting coherence length $\xi$$\sim$20\AA, and the 218\AA\ distance 
between vortices (at $H$=5T). The large value of $\zeta$ makes it unlikely 
that the antiferromagnetic signal arises from individual vortex cores, which 
have spatial extent $\xi$. The onset of the signal at a temperature in 
excess of $T_{\rm irr}$($H$=5T) is incompatible with interpretation of the
irreversibility line as a vortex melting line, because vortex motion 
would dephase the IC order nucleated by the vortices. A more plausible
account of our data proceeds from the hypothesis that the cores nucleate
larger, immobile IC regions, which overlap both with each other and the
pre-existing IC background to increase the IC volume fraction. Such
agglomeration would occur as soon as the phase separation proposed by the
theory of type II superconductors becomes advantageous, and leaves only
small interstitial regions with SC phase coherence. Phase coherent
superconductivity would then be established throughout the sample at the
lower 'irreversibility' temperature via Josephson coupling between the
finite-sized SC regions, rather than via freezing of a liquid of mobile
vortices. 

\begin{figure}
\caption{
Magnetotransport and neutron diffraction data for La$_{2-x}$Sr$_{x}$CuO$_{4}$ 
as a function of temperature and magnetic field. Our sample was grown 
in an optical image furnace and has strontium content $x$=0.10, putting 
it in the underdoped superconducting regime with $T_{c}$($H$=0T)=29K.
a) shows magnetotransport measurements parallel to the CuO$_{2}$ planes, 
obtained via a standard four-probe method; the colours indicate the electrical 
resistivity. In a magnetic field, the sharp transition from normal to 
superconducting states is broadened into a crossover region and vortices are 
thought to form at temperatures where the resistivity falls below its value 
at $T_{c}$($H$=0). Phase coherent superconductivity, characterised by zero 
resistance, sets in at the much lower ('irreversibility') temperature, 
$T_{\rm irr}$($H$), marked by the white circles.
b) shows the temperature-dependent intensity (above background) of the 
neutron diffraction signal for zero field (blue circles) and a field of 
5T applied perpendicular to the CuO$_{2}$ layers (red squares). The lines 
are guides to the eye. Measurements were performed on a 1.5cm$^{3}$ sample, 
using the V2 triple-axis spectrometer at the Hahn-Meitner Institute, Berlin. 
The inset shows the relevant reciprocal space, labeled using the 
two-dimensional notation appropriate for the superconducting planes. The black 
circle at ($\frac{1}{2}$,$\frac{1}{2}$) represents the Bragg point associated 
with the commensurate antiferromagnetism of the insulating $x$=0 parent 
compound. The incommensurate antiferromagnetic order in our metallic 
$x$=0.10 material, gives rise to diffraction at the quartet of red circles 
(($\frac{1}{2}$,$\frac{1}{2}$)+(0,$\pm$0.12) and 
($\frac{1}{2}$,$\frac{1}{2}$)+($\pm$0.12,0)) where our measurements took 
place. 
}
\end{figure}
\epsfbox{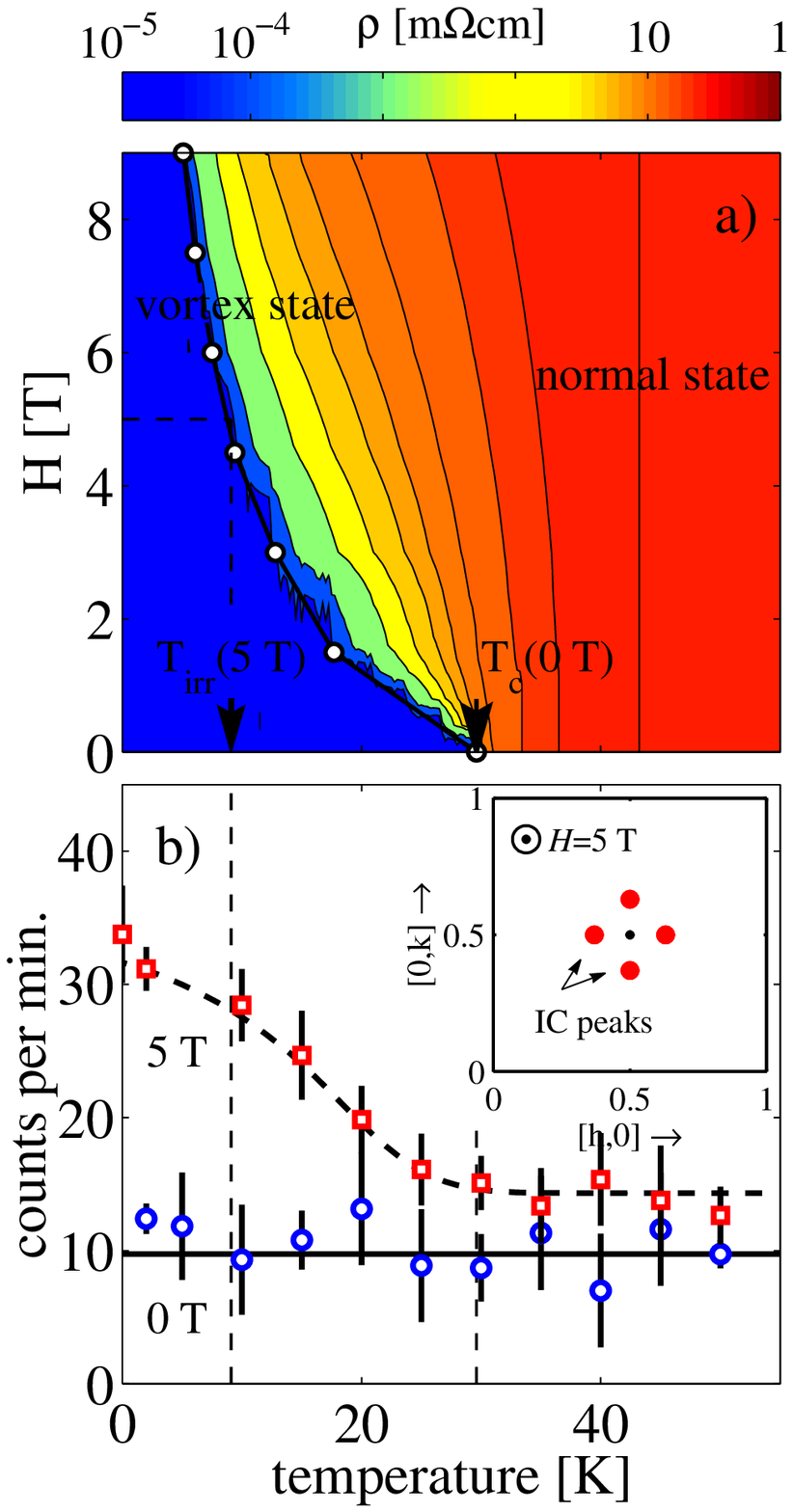}
\end{document}